\documentclass[aps,pre,reprint,groupedaddress,amsmath,amssymb]{revtex4-1}

\usepackage{bm}
\usepackage{graphicx}
\usepackage{color}

\begin{document}


\title{Delayed feedback control of synchronization in weakly coupled oscillator networks}


\author{Viktor Novi\v{c}enko}
\email[]{novicenko@pfi.lt}
\homepage[]{http://www.itpa.lt/~novicenko/}
\affiliation{Institute of Theoretical Physics and Astronomy, Vilnius University, A. Go\v{s}tauto 12, LT-01108 Vilnius, Lithuania}


\date{\today}

\begin{abstract}
We study control of synchronization in weakly coupled oscillator networks by using a phase reduction approach. Starting from a general class of limit cycle oscillators we derive a phase model, which shows that delayed feedback control changes effective coupling strengths and effective frequencies. We derive the analytical condition for critical control gain, where the phase dynamics of the oscillator becomes extremely sensitive to any perturbations. As a result the network can attain phase synchronization even if the natural interoscillatory couplings are small. In addition, we demonstrate that delayed feedback control can disrupt the coherent phase dynamic in synchronized networks. The validity of our results is illustrated on networks of diffusively coupled Stuart--Landau and FitzHugh--Nagumo models.
\end{abstract}


\maketitle


\section{Introduction}

Starting from C.~Huygens' research on ``an odd kind sympathy'' between coupled pendulum clocks, the synchronization as a phenomenon occurs in various man-made and natural systems~\cite{kura03,winf01,pikov01,izhi07}. The coherent behavior of oscillators arises in numerous situations, \textit{e.g.}, flashing of fireflies~\cite{Buck1988}, cardiac pacemaker cells~\cite{Glass1988}, neurons in the brain~\cite{Varela2001}, coupled Josephson junctions~\cite{PhysRevE.57.1563}, chemical reactions~\cite{kura03,Kiss2002}, crowd synchrony~\cite{Strogatz2005}, and power grids~\cite{Motter2013,Dorfler2013}. The synchronous behavior can be desirable or harmful. The ability to control synchrony in oscillatory networks covers a wide range of real-world applications, starting from neurological treatment of Parkinson's disease and essential tremor~\cite{benabid1991,benabid2002} to the design of robust power grids~\cite{Dorfler2013,Dorfler2014}.

Phase reduction is a fundamental theoretical technique to investigate synchronization in weakly coupled oscillator networks~\cite{kura03,winf01,pikov01,izhi07}, since it allows the approximation of high-dimensional dynamics of oscillators with a single phase variable. The concept of the phase model causes significant progress in understanding the synchrony of the networks, \textit{e.g.}, correlation between topology and dynamics towards synchronization~\cite{PhysRevLett.96.114102}, synchronization criterion for almost any network topology~\cite{Dorfler2013}, optimal synchronization~\cite{Skardal2014}, chimera states~\cite{kuramoto2002,PhysRevLett.93.174102}, \textit{etc}. The main factors determining the synchrony in the phase model are coupling strength and dissimilarity of frequencies. The ability to change these parameters will easily allow the synchronization or desynchronization of networks. Typically, the phase variable is not attained for direct measurements and actions. Instead of this, we have an access to dynamical variables of the limit cycle. In such situations, the control schemes are usually based on feedback loops. Therefore, we ask: how do we enhance or suppress synchronization in networks via feedback signals, when minimal knowledge about the particular unit of the network is available? This question has been investigated in Refs.~\cite{ros04,PhysRevE.70.041904}, where the synchronization is controlled by delayed mean-field feedback into the network. Also, there have been many numerical investigations~\cite{Brandstetter2009, Hovel2010, hov_conf2007, Schoell2009} devoted to this question. In this work, we present analytical results for control of synchronization in networks by time delayed local signals fed back into particular units of the network. By using phase reduction for systems with time-delay~\cite{physd12,kot12}, we arrive at a phase model. It fully coincides with the phase model of the uncontrolled network, the only difference being that the coupling strengths and frequencies depend on the control parameters. Surprisingly, the relations are almost universal, \textit{i.e.} do not depend on the particular model of the limit cycle and on the coupling interoscillatory function. Moreover, the coupling strengths have a multiplicative inverse dependence on the feedback control gain and their values can be selected from zero to infinity. As a consequence, the synchronization can be achieved even if the oscillators are almost uncoupled. Also, we show that the particular choice of the delay times in the control scheme can lead to full phase synchronization (\textit{i.e.}, when the phases of all oscillators are equal at any time moment) in the network. The analytical results are verified numerically on networks of diffusively coupled Stuart--Landau and FitzHugh--Nagumo models.

\section{Phase reduction of oscillator network}

We consider a general class of $N$ weakly coupled limit cycle oscillators under delayed feedback control~(DFC)
\begin{equation}
\dot{\mathbf{x}}_i = \mathbf{f}_i\left(\mathbf{x}_i \right)+\varepsilon \sum_{j=1}^N \mathbf{G}_{ij}\left(\mathbf{x}_i,\mathbf{x}_j \right)+\mathbf{K}_i \left[\mathbf{x}_i(t-\tau_i)-\mathbf{x}_i(t) \right],
\label{main}
\end{equation}
where $\mathbf{x}_i \in \mathbb{R}^n$ is an $n$-dimensional state vector of the $i$th oscillator, $\mathbf{f}_{i}: \mathbb{R}^n \rightarrow \mathbb{R}^n$ is a vector field representing the free dynamics of the $i$th oscillator, $\mathbf{G}_{ij}: \mathbb{R}^n \times \mathbb{R}^n \rightarrow \mathbb{R}^n$ is an inter-oscillatory coupling function, $\mathbf{K}_i=\textrm{diag}[K_i^{(1)},K_i^{(2)},\ldots,K_i^{(n)}]$ is an $n$-dimensional diagonal matrix of the feedback control gain, and $\tau_i$ is the delay time of the $i$th oscillator's feedback loop. We assume that $\varepsilon>0$ is a small parameter. Each uncoupled oscillator $\dot{\mathbf{x}}_i = \mathbf{f}_i\left(\mathbf{x}_i \right)$ has the stable  limit-cycle solution $\bm{\xi}_i(t+T_i)=\bm{\xi}_i(t)$ with the natural frequency $\Omega_i=2\pi/T_i$. We are interested in the case when dissimilarity of the periods is of the order of $\varepsilon$, i.e., $(T_i-T_j) \sim \varepsilon$ for any $i,j=1,\ldots, N$. Also, we assume that the delay times are close to the periods, i.e., $(T_i-\tau_i) \sim \varepsilon$.

By expanding delayed vector $\mathbf{x}_i(t-\tau_i)$ into a Taylor series and omitting higher-than-$\varepsilon$-order terms, we arrive at the following expression for the control force:
\begin{eqnarray}
& & \mathbf{K}_i \left[\mathbf{x}_i(t-\tau_i)-\mathbf{x}_i(t) \right] \approx  \nonumber \\ & & \mathbf{K}_i \left[\mathbf{x}_i(t-T_i)-\mathbf{x}_i(t) \right]-\mathbf{K}_i \dot{\mathbf{x}}_i(t-T_i) \Delta T_i,
\label{contr_f}
\end{eqnarray}
where $\Delta T_i=\tau_i-T_i$ is the mismatch of the time-delay. The first term on the right hand side~(r.h.s.) of expression (\ref{contr_f}) is familiar from the controlling chaos, where it is used to stabilize unstable periodic orbits in chaotic systems~\cite{pyr92,pyr06}. Therefore, we use well-known results such as the odd number limitation theorem~\cite{hoo12} and mismatched control scheme~\cite{nov12}.

The oscillators under DFC,
\begin{equation}
\dot{\mathbf{x}}_i = \mathbf{f}_i\left(\mathbf{x}_i \right)+\mathbf{K}_i \left[\mathbf{x}_i(t-T_i)-\mathbf{x}_i(t) \right]
\label{osc_dfc}
\end{equation}
also have the same periodic solutions $\bm{\xi}_i(t)$ as the free oscillators, but with different stability properties, and as a consequence, with different perturbation-induced phase response. If the free oscillator has an infinitesimal phase-response curve~(iPRC) $\mathbf{z}_i(\vartheta_i)$ [the iPRC is a $T_i$-periodic solution of the adjoint equation $\dot{\mathbf{z}}_i(t)=-[D \mathbf{f}_i(\bm{\xi}_i(t))]^T\mathbf{z}_i(t)$ with the initial condition $\mathbf{z}_i^T(0)\cdot\dot{\bm{\xi}}_i(0)=1$; see \cite{kura03,winf01,pikov01,izhi07}] then the oscillator under DFC~(\ref{osc_dfc}) has the iPRC $\mathbf{z}_{i}^{(\textrm{DFC})}(\vartheta_i)$ of the same form but with different amplitude~\cite{physd12}: $\mathbf{z}_{i}^{(\textrm{DFC})}(\vartheta_i)=\alpha_i \mathbf{z}_i(\vartheta_i)$. The factor $\alpha_i$ can be expressed as:
\begin{equation}
\alpha_i=\alpha_i(\mathbf{K}_i)=\left[ 1+\sum_{m=1}^n K_i^{(m)}C_i^{(m)} \right]^{-1}.
\label{alpha}
\end{equation}
Here the coefficients $C_i^{(m)}$ are the following integrals: $C_i^{(m)}=\int_0^{T_i} z_i^{(m)}(s)\dot{\xi}_i^{(m)}(s) d s$, where the upper indices $^{(m)}$ denote the particular components of the vectors $\mathbf{z}_i$ and $\dot{\bm{\xi}}_i$.

Now we apply the phase reduction technique~\cite{physd12,kot12} to the oscillator network (\ref{main}) assuming that the unperturbed oscillators are described by equations (\ref{osc_dfc}) and that the perturbation contains two parts: the interoscillatory coupling terms $\mathbf{G}_{ij}$ and the second term on the r.h.s. of expression (\ref{contr_f}). Both parts are of the same order: $O(\varepsilon)$. The equations for the phase dynamics are
\begin{eqnarray}
\dot{\vartheta}_i &=& 1+\varepsilon\alpha_i(\mathbf{K}_i)\mathbf{z}_i^T(\vartheta_i)\sum_{j=1}^N \mathbf{G}_{ij}\left(\bm{\xi}_i(\vartheta_i),\bm{\xi}_j(\vartheta_j) \right) \nonumber \\
& & -\Delta T_i \alpha_i(\mathbf{K}_i)\mathbf{z}_i^T(\vartheta_i)\mathbf{K}_i \dot{\bm{\xi}}_i(\vartheta_i).
\label{phase}
\end{eqnarray}
Here in the last term of the r.h.s. we write $\dot{\bm{\xi}}_i(\vartheta_i(t))$ instead of $\dot{\bm{\xi}}_i(\vartheta_i(t-T_i))$. It can be done because
\begin{equation}
\dot{\bm{\xi}}_i(\vartheta_i(t-T_i)) = \dot{\bm{\xi}}_i(\vartheta_i(t)+O(\varepsilon)) = \dot{\bm{\xi}}_i(\vartheta_i(t))+O(\varepsilon)
\label{corr}
\end{equation}
and after multiplication by $\Delta T_i$ we get the second order correction $O(\varepsilon^2)$ which is omitted~\cite{nov12}.

The equations for the phase dynamics (\ref{phase}) are valid only when all periodic solutions $\bm{\xi}_i(t)$ are stable solutions of the system~(\ref{osc_dfc}). According to the odd number limitation theorem~\cite{hoo12}, the periodic solution $\bm{\xi}_i(t)$ is the unstable solution of the system~(\ref{osc_dfc}) if the condition
\begin{equation}
\sum_{m=1}^n K_i^{(m)}C_i^{(m)}<-1
\label{onl}
\end{equation}
holds. The condition~(\ref{onl}) shows which values of the feedback control gains cannot be correctly described by Eq.~(\ref{phase}). If this condition does not hold, then it is still not guaranteed that the solution $\bm{\xi}_i(t)$ is stable and that Eq.~(\ref{phase}) is valid. However, as we see below, for particular systems (namely, diffusively-coupled Stuart--Landau and FitzHugh--Nagumo models) the condition~(\ref{onl}) is necessary and sufficient.

The phases $\vartheta_i(t)$ in equation~(\ref{phase}) vary from $0$ to $T_i$. However, when we investigate phase synchronization, it is convenient to have phases growing from $0$ to $2\pi$. Furthermore, on the r.h.s. of Eq.~(\ref{phase}), the first term corresponds to trivial growing of the phase. Therefore, we introduce new variables $\varphi_i(t)=\Omega_i\vartheta_i(t)-\frac{2 \pi}{T}t$, where $T$ is the so-called ``averaged'' period. The number $T$ is not necessarily equal to the average of all oscillator periods and can be chosen freely with one requirement: $(T-T_i) \sim \varepsilon$ for all $i=1,\ldots, N$. In the new variables, Eq.~(\ref{phase}) can be written as
\begin{eqnarray}
\dot{\varphi}_i &=& \omega_i+\varepsilon\Omega_i\alpha_i(\mathbf{K}_i)\mathbf{z}_i^T\left(\frac{\varphi_i}{\Omega_i}+ \frac{\Omega}{\Omega_i} t \right)\nonumber \\
& & \times \sum_{j=1}^N \mathbf{G}_{ij}\left(\bm{\xi}_i\left(\frac{\varphi_i}{\Omega_i}+ \frac{\Omega}{\Omega_i} t \right),\bm{\xi}_j\left(\frac{\varphi_j}{\Omega_j}+ \frac{\Omega}{\Omega_j} t \right) \right) \nonumber \\
& & -\Delta T_i \Omega_i \alpha_i(\mathbf{K}_i)\mathbf{z}_i^T\left(\frac{\varphi_i}{\Omega_i}+ \frac{\Omega}{\Omega_i} t \right)\mathbf{K}_i \dot{\bm{\xi}}_i\left(\frac{\varphi_i}{\Omega_i}+ \frac{\Omega}{\Omega_i} t \right), \nonumber
\label{phase1}
\end{eqnarray}
where $\Omega=2\pi/T$ is the ``averaged'' frequency and $\omega_i=\Omega_i-\Omega$. The r.h.s. of the last equation depends periodically on time with period $T$; also, $\omega_i$ and $\Delta T_i$ are small parameters. Thus we can apply the averaging method~\cite{burd07,sand07}. Let us denote the averaged phases $\psi_i(t)$. The phase model for the averaged phases is
\begin{equation}
\dot{\psi}_i = \omega_i^{\textrm{eff}}+\varepsilon_i^{\textrm{eff}} \sum_{j=1}^N H_{ij}(\psi_j-\psi_i),
\label{phase_avr}
\end{equation}
where we introduce effective coupling strengths
\begin{equation}
\varepsilon_i^{\textrm{eff}} = \varepsilon \alpha_i(\mathbf{K}_i),
\label{eff_e}
\end{equation}
effective frequencies
\begin{equation}
\omega_i^{\textrm{eff}} = \omega_i+\Omega\frac{\Delta T_i}{T}\left[\alpha_i(\mathbf{K}_i)-1 \right],
\label{eff_freq}
\end{equation}
[in equation (\ref{eff_freq}) the index $i$ near the $\Omega$ and $T$ is skipped without loss of accuracy] and coupling functions
\begin{eqnarray}
& & H_{ij}(\chi) = \frac{1}{T_i} \nonumber \\
& & \times \int_0^{2\pi}\mathbf{z}_i^T\left(\frac{s}{\Omega_i} \right) \mathbf{G}_{ij}\left(\bm{\xi}_i\left(\frac{s}{\Omega_i}\right),\bm{\xi}_j\left(\frac{\chi+s}{\Omega_j} \right) \right) d s.
\label{coup_fun}
\end{eqnarray}

Hereafter we assume that all network units are near identical and described by similar equations, i.e., $[\mathbf{f}_i(\bm{\xi}_i(t))-\mathbf{f}_j(\bm{\xi}_j(t))] \sim \varepsilon$ for all $i,j=1\ldots N$. We denote ``averaged'' oscillator as $\dot{\mathbf{x}} = \mathbf{f}\left(\mathbf{x} \right)$, which has stable periodic solution $\bm{\xi}(t+T)=\bm{\xi}(t)$ and the corresponding iPRC $\mathbf{z}(t+T)=\mathbf{z}(t)$. The choice of the ``averaged'' oscillator must satisfy one requirement:  $[\mathbf{f}(\bm{\xi}(t))-\mathbf{f}_i(\bm{\xi}_i(t))] \sim \varepsilon$ for all $i=1\ldots N$. Also, we assume that interoscillatory functions have the form $\mathbf{G}_{ij}\left(\mathbf{x}_i,\mathbf{x}_j \right)=a_{ij} \mathbf{g} \left(\mathbf{x}_i,\mathbf{x}_j \right)$ where the coefficients $a_{ij}\geq0$ play the role of the network's adjacency matrix elements. We consider the undirected network, therefore the adjacency matrix $\mathbf{A}^T=\mathbf{A}$. Now we can simplify the network's phase model (\ref{phase_avr}). Since $\bm{\xi}_i(s/\Omega_i)=\bm{\xi}(s/\Omega)+O(\varepsilon)$ and $\mathbf{z}_i(s/\Omega_i)=\mathbf{z}(s/\Omega)+O(\varepsilon)$, without loss of accuracy, the indices $i$ near $\alpha$ can be dropped and Eq.~(\ref{phase_avr}) can be rewritten as
\begin{equation}
\dot{\psi}_i = \omega_i^{\textrm{eff}}+\varepsilon_i^{\textrm{eff}} \sum_{j=1}^N a_{ij} h(\psi_j-\psi_i),
\label{phase_avr1}
\end{equation}
with effective coupling strengths $\varepsilon_i^{\textrm{eff}} = \varepsilon \alpha(\mathbf{K}_i)$, effective frequencies $\omega_i^{\textrm{eff}} = \omega_i+\Omega\left(\Delta T_i/T\right)\left[\alpha(\mathbf{K}_i)-1 \right]$, coupling functions
\begin{eqnarray}
& & h(\chi) = \frac{1}{T} \nonumber \\
& & \times \int_0^{2\pi}\mathbf{z}^T\left(\frac{s}{\Omega} \right) \mathbf{g}\left(\bm{\xi}\left(\frac{s}{\Omega}\right),\bm{\xi}\left(\frac{\chi+s}{\Omega} \right) \right) d s,
\label{coup_fun1}
\end{eqnarray}
and the factor $\alpha(\mathbf{K}_i)=\left[1+\sum_{m=1}^n K_i^{(m)}C^{(m)} \right]^{-1}$, where the integrals $C^{(m)}=\int_0^{T} z^{(m)}(s)\dot{\xi}^{(m)}(s) d s$.

Our main result is the phase model~(\ref{phase_avr1}). Using (\ref{phase_avr1}), we can analyze three important control regimes. Before that let us make two additional assumptions: 1) - we assume that the condition (\ref{onl}) is necessary and sufficient, \textit{i.e.}, the periodic solutions $\bm{\xi}_i(t)$ are stable till~(\ref{onl}) does not hold, and 2) - we assume that for uncontrolled network ($\mathbf{K}_i=0$) there exist a positive threshold coupling strength $\varepsilon_{\textrm{th}}>0$ such that when $\varepsilon>\varepsilon_{\textrm{th}}$ the network possess stable phase synchronization regime
\begin{equation}
\dot{\psi}_1=\dot{\psi}_2=\ldots = \dot{\psi}_N,
\label{sync}
\end{equation}
while for $\varepsilon<\varepsilon_{\textrm{th}}$ the phase synchronization cannot be achieved. Hence the important control cases are

(i) If all delay times and feedback control gains are equal ($\tau_i=\tau$ and $\mathbf{K}_i=\mathbf{K}$ for all $i=1\ldots N$), then synchronization of the network cannot be controlled. In this case the phase model~(\ref{phase_avr1}) is equivalent to phase model of uncontrolled network. It can be seen, if we choose ``averaged'' period $T=\tau$ (without loss of generality we always can do that) and rewrite effective frequencies
\begin{eqnarray}
\omega_i^{\textrm{eff}} & = & \omega_i+\Omega\frac{\Delta T_i}{T}\left[\alpha(\mathbf{K})-1 \right] \nonumber \\
& = &\omega_i+\omega_i \left[\alpha(\mathbf{K})-1 \right]+O(\varepsilon^2)\approx \omega_i\alpha(\mathbf{K}).
\label{eff_freq_rew}
\end{eqnarray}
Since $\varepsilon_i^{\textrm{eff}} = \varepsilon \alpha(\mathbf{K})$, the factor $\alpha(\mathbf{K})$ can be eliminated from Eqs.~(\ref{phase_avr1}) by time-scaling transformation.

(ii) If all delay times are equal to the periods ($\tau_i=T_i$ for all $i=1\ldots N$), then $\omega_i^{\textrm{eff}} = \omega_i$, and it is possible to synchronize or desynchronize network independently on how big or small the natural interoscillatory coupling is. Let us say that all diagonal matrices $\mathbf{K}_i$ have only one nonzero element $K_i^{(1)}=K^{(1)}$ and assume that $C^{(1)}$ is positive. Then, from condition (\ref{onl}) the phase model~(\ref{phase_avr1}) is valid, if $K^{(1)}$ is in the interval $(-1/C^{(1)},+\infty)$. The effective coupling strengths $\varepsilon_i^{\textrm{eff}} = \varepsilon/[1+K^{(1)}C^{(1)}]$ go to infinity if $K^{(1)}\rightarrow -1/C^{(1)} + 0$. Therefore, the oscillators' phases become extremely sensitive to any perturbations. At boundary $K^{(1)}=-1/C^{(1)}$ all oscillators become neutrally stable. For the other boundary, if $K^{(1)}\rightarrow +\infty$ then $\varepsilon_i^{\textrm{eff}} \rightarrow 0$, and the oscillators' phases cannot synchronize to each other. Note that even if a magnitude of the effective coupling strength can be chosen freely, the sign cannot be changed.

(iii) If the mismatch times are equal to
\begin{equation}
\frac{\Delta T_i}{T} = \frac{\omega_i}{\Omega \left[1-\alpha(\mathbf{K}_i) \right]},
\label{fps}
\end{equation}
then $\omega_i^{\textrm{eff}} = 0$. In this case the network possesses stable full phase synchronization regime
\begin{equation}
\psi_1(t)=\psi_2(t)=\ldots = \psi_N(t)=\textrm{const}
\label{full_phase_sync}
\end{equation}
under additional conditions: $h(0)=0$ and $h'(0)=\gamma>0$. The condition $h(0)=0$ is always fulfilled if the units are diffusively-coupled, and the condition $\gamma>0$ represents attractive coupling. The stability of the full phase synchronization regime (\ref{full_phase_sync}) is determined by eigenvalues of a matrix $\mathbf{M}=-\gamma \mathbf{E} \mathbf{L}$, where $\mathbf{E}=\textrm{diag}[\varepsilon_1^{\textrm{eff}},\varepsilon_2^{\textrm{eff}},\ldots,\varepsilon_N^{\textrm{eff}}]$ is a diagonal positive-definite matrix and $\mathbf{L}=\mathbf{D}-\mathbf{A}$ is a network Laplacian matrix (here $\mathbf{D}=\textrm{diag}[d_1,d_2,\ldots,d_N]$ is the degree matrix with the elements $d_i=\sum_{j=1}^N a_{ij}$). We assume that the network is connected and unidirected, so the matrix $\mathbf{L}$ has eigenvalues $0=\lambda_1<\lambda_2\leq \ldots \leq \lambda_N$. By defining a square root of the matrix $\mathbf{E}$ as $\mathbf{E}^{1/2}$ with the entries $(\varepsilon_i^{\textrm{eff}})^{1/2}$ on the diagonal, we can see that $\mathbf{M}$ has the same eigenvalues as a symmetric matrix $\mathbf{M}'=-\gamma \mathbf{E}^{1/2} \mathbf{L} \mathbf{E}^{1/2}$. $\mathbf{M}'$ is negative semidefinite matrix with only one eigenvalue equal to zero, which corresponds to a shift of all phases by a same amount. Hence the full phase synchronization regime is stable.

\section{Numerical demonstrations}

\begin{figure}[t!]
\centering\includegraphics[width=0.99\columnwidth]{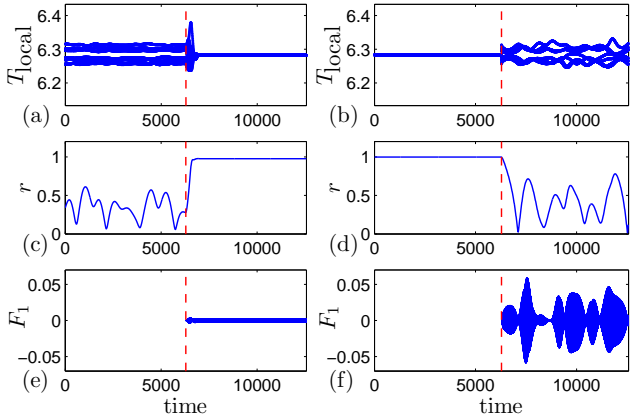}
\caption{\label{fig1} (Color online) Control of synchronization in the SL network~(\ref{sl}) for the mismatches $\Delta T_i=0$. (a) and (b) ``local'' periods, (c) and (d) Kuramoto order parameter, (e) and (f) DFC force applied to the first unit $F_1(t)=K\left[x_1(t-T_1)-x_1(t) \right]$. The vertical dashed red line shows time moment, when the control is turned-on. (a), (c), (e) DFC causes synchronization in network with parameters $\varepsilon=9\times 10^{-4}$ and $K=-0.3$, since $\varepsilon^{\textrm{eff}}>\varepsilon_{\textrm{th}}$. (b), (d), (f) DFC causes desynchronization in network with parameters $\varepsilon=5 \times 10^{-2}$ and $K=4$, since $\varepsilon^{\textrm{eff}}<\varepsilon_{\textrm{th}}$.}
\end{figure}
\begin{figure}[t!]
\centering\includegraphics[width=0.95\columnwidth]{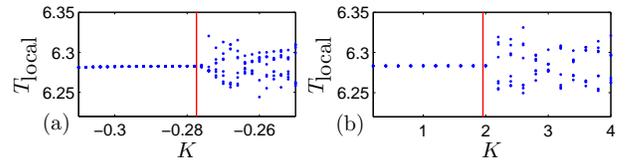}
\caption{\label{fig1_a} (Color online) The synchrony-asynchrony transition in the SL network for the control regime $\Delta T_i = 0$. A snapshot of the ``local'' periods versus the control gain calculated numerically for different coupling strengths: (a) $\varepsilon=9\times 10^{-4}$ and (b) $\varepsilon=5 \times 10^{-2}$. A vertical red line shows analytically derived critical value of the control gain.}
\end{figure}

As a first example, we study a network of $N=8$ all-to-all diffusively coupled Stuart--Landau (SL) models described by the following equations:
\begin{subequations}
\label{sl}
\begin{eqnarray}
\dot{x}_i &=& x_i\left(1-x_i^2-y_i^2 \right) -\Omega_i y_i+\varepsilon N^{-1} \sum_{j=1}^N 2(x_j-x_i) \nonumber \\ 
& & +K \left[ x_i(t-\tau_i) - x_i(t)\right], \label{sl_1} \\
\dot{y}_i &=& y_i\left(1-x_i^2-y_i^2 \right) +\Omega_i x_i. \label{sl_2}
\end{eqnarray}
\end{subequations}
As an ``averaged'' oscillator we choose SL model with $\Omega=1$. For this case, the periodic orbit and iPRC can be found analytically: $\bm{\xi}(t)=\left[ \cos t, \sin t \right]^T$ and $\mathbf{z}(t)=\left[ -\sin t, \cos t \right]^T$. According to (\ref{coup_fun1}), the coupling function $h(\chi)=\sin \chi$. For the oscillators natural frequencies we choose following values $\Omega_i=\Omega+\omega_i$, where $\omega_i=10^{-3}\times \left\lbrace 1.38, \, 2.54, \, -1.93, \, -4.87, \, -2.12, \, 3.95, \, 4.31, \, -3.26 \right\rbrace$. For SL model the factor $\alpha(K)=\left[1+K \pi \right]^{-1}$ and the feedback control gain $K$ can be selected from the interval $(-\pi^{-1}, \, +\infty)$. We calculate numerically that the network~(\ref{sl}) without control ($K=0$) possesses phase synchronization if $\varepsilon$ is above the threshold value $\varepsilon_{\textrm{th}}=7\times 10^{-3}$. The adjacency matrix of the network~(\ref{sl}) is $a_{ij}=N^{-1}$ and the corresponding phase model is the celebrated Kuramoto model
\begin{equation}
\dot{\psi}_i = \omega_i^{\textrm{eff}}+\varepsilon^{\textrm{eff}}N^{-1} \sum_{j=1}^N \sin (\psi_j-\psi_i).
\label{sl_ph}
\end{equation}
As a synchronization criteria we choose two measurements: Kuramoto order parameter $r(t)=N^{-1} \left| \sum_{j=1}^N \exp(i\psi_j(t))\right|$ and the ``local'' periods $T_{\textrm{local}}$ (or sometimes called interspike intervals) defined as a time interval between two neighboring maxima of the first dynamical variable. Figure~\ref{fig1} shows numerical simulation of the SL network~(\ref{sl}) when the mismatches $\Delta T_i=0$. The transition synchrony-asynchrony occurs at a critical control gain $K_c=\pi^{-1}\left[\varepsilon/\varepsilon_{\textrm{th}} -1 \right]$. In Figure.~\ref{fig1_a} we demonstrate the synchrony-asynchrony transition in the SL network. As we can see, analytical results coincide with numerical simulations.

\begin{figure}[h!]
\centering\includegraphics[width=0.9\columnwidth]{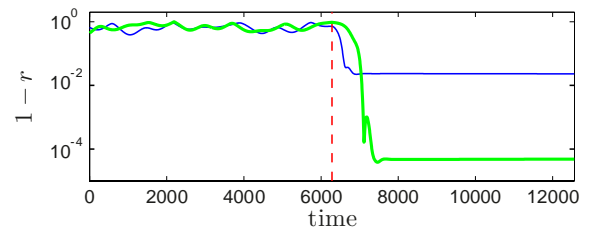}
\caption{\label{fig2} (Color online) Kuramoto order parameters of SL network (\ref{sl}) depicted on a semilog plot. The blue thin line reproduces results from Fig.~\ref{fig1}~(c), while the green thick line calculated in full-phase-synchronization regime. The vertical dashed red line shows time moment, when the control is turned-on.}
\end{figure}

In order to demonstrate full phase synchronization regime, we simulate SL network with the same parameters as presented in Figs.~\ref{fig1} (a), (c) and (e), only the mismatch times are selected according to (\ref{fps}). The ``local'' periods and $F_1$ are very similar to that presented in Figs.~\ref{fig1} (a) and (e), only the Kuramoto order parameter is much closer to one (cf. Fig.~\ref{fig2}).

\begin{figure}[t!]
\centering\includegraphics[width=0.40\columnwidth]{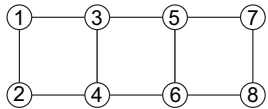}
\caption{\label{fig3} Topology of the FitzHugh-Nagumo oscillator network~(\ref{fhn}).}
\end{figure}

\begin{figure}[t!]
\centering\includegraphics[width=0.90\columnwidth]{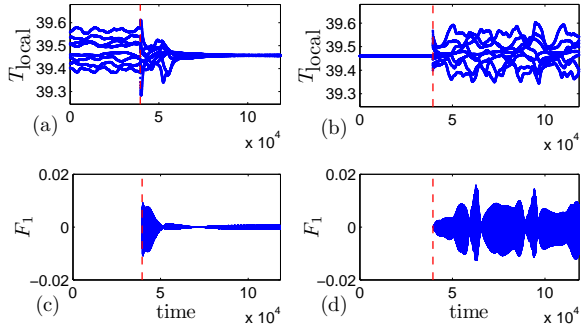}
\caption{\label{fig4} (Color online) Control of synchronization in the FHN network~(\ref{fhn}) for the mismatches $\Delta T_i=0$. (a) and (b) ``local'' periods, (c) and (d) DFC force applied to the first unit $F_1(t)=K\left[x_1(t-T_1)-x_1(t) \right]$. The vertical dashed red line shows time moment, when the control is turned-on. (a), (c) DFC causes synchronization in network with parameters $\varepsilon=5\times 10^{-5}$ and $K=-0.09$, since $\varepsilon^{\textrm{eff}}>\varepsilon_{\textrm{th}}$. (b), (d) DFC causes desynchronization in network with parameters $\varepsilon= 10^{-3}$ and $K=0.5$, since $\varepsilon^{\textrm{eff}}<\varepsilon_{\textrm{th}}$.}
\end{figure}

In order to show that analytical results valid for nontrivial oscillators and for the nontrivial network topology, we investigate the network of $N=8$ diffusivelycoupled FitzHugh--Nagumo (FHN)~\cite{Fitzhugh61, Nag62} models
\begin{subequations}
\label{fhn}
\begin{eqnarray}
\dot{x}_i &=& x_i-x_i^3/3-y_i+0.5+\varepsilon\sum_{j=1}^N a_{ij}(x_j-x_i) \nonumber \\ 
& & +K \left[ x_i(t-\tau_i) - x_i(t)\right], \label{fhn_1} \\
\dot{y}_i &=& \epsilon_i (x_i+0.7-0.8y_i). \label{fhn_2}
\end{eqnarray}
\end{subequations}
The adjacency matrix elements $a_{ij}=1$, if the unit $i$ is connected to the unit $j$, and $a_{ij}=0$ otherwise. The network topology is illustrated in Fig.~\ref{fig3}.
As an ``averaged'' oscillator we choose FHN model with $\epsilon=0.08$. For such model the constant $C^{(1)} \approx 10.02$ computed numerically. We check that ``averaged'' oscillator possesses stable periodic solution $\bm{\xi}(t)$, when control gain $K$ is in the  interval $(-1/C^{(1)},+\infty)$. In the network (\ref{fhn}) each oscillator has different parameter $\epsilon_i=\epsilon+\Delta\epsilon_i$, where $\Delta\epsilon_i=10^{-4}\times \left\lbrace 0.3, \, -1.7, \, -0.9, \, 2.1, \, 1.5, \, -2.6, \, -1.1, \, 0.8 \right\rbrace$, and without control ($K=0$) it possesses phase synchronization if $\varepsilon$ is above the threshold $\varepsilon_{\textrm{th}}=3.6 \times 10^{-4}$. Figure~\ref{fig4} shows numerical simulation of the FHN network~(\ref{fhn}) when the mismatches $\Delta T_i=0$. Again, analytical results coincide with numerical simulations.

\section{Conclusion}

We present framework for controlling synchrony in weakly coupled oscillator networks by delayed feedback control. We show that when the delay time is close to the period of a particular oscillator, the network's phase model almost coincides with the uncontrolled network's phase model. The only difference is that effective coupling strengths and effective frequencies depend on control parameters. By appropriate choice of the control parameters the magnitude of the effective coupling strength can be selected arbitrary, while the sign cannot be changed. Unlike coupling strength, the sign of the effective frequencies can be inverted.

In this work we have restricted ourselves to the case when control term appears as an external force applied to the oscillator. However it can be simply generalized to the case of arbitrary functional dependence of the oscillator on the control signal.

\begin{acknowledgments}
Author would like to thank Vaidas Juknevi\v{c}ius, Julius Ruseckas and Art\={u}ras Novi\v{c}enko for careful reading and correcting the manuscript of this paper, and Irmantas Ratas for fruitful discussions.
\end{acknowledgments}

\bibliography{references}

\end{document}